\documentstyle[12pt,epsf]{article}
  \newcommand{\N}{{\rm I \hspace{-0.52ex} N}}
\begin{document}
\setlength\textheight{8.75in}
\newcommand{\be}{\begin{equation}}
\newcommand{\ee}{\end{equation}}
\title{On PT-symmetric extensions of the Calogero and Sutherland
models}
\author{{\large Yves Brihaye\footnote{ yves.brihaye@umh.ac.be}} \\
{\small Facult\'e des Sciences, Universit\'e de Mons-Hainaut, }\\
{\small B-7000 Mons, Belgium }\\
{ } \\
   {\large Ancilla Nininahazwe
   \footnote{nininaha@yahoo.fr}}\\
{\small   Facult\'e des Sciences,
Universit\'e du Burundi}\\
{\small P.O. Box 2700, Bujumbura, Burundi}}

\date{\today}
\maketitle
\thispagestyle{empty}

\begin{abstract}
The original Calogero and Sutherland models describe N 
quantum particles
on the line interacting pairwise through an inverse square
and an inverse sinus-square potential.
They  are well known to be integrable and solvable.
Here we extend the Calogero and Sutherland Hamiltonians 
by means of
new interactions which are PT-symmetric but not self adjoint.
Some of these new interactions lead to
integrable PT-symmetric Hamiltonians; the algebraic properties
further reveal that they are solvable as well.
We also consider  PT-symmetric interactions which
lead to a new quasi-exactly
solvable deformation of the Calogero and 
Sutherland Hamiltonians.
\end{abstract}
\medskip
\medskip
\newpage
\section{Introduction}

The Calogero and Sutherland models \cite{calogero,suth} 
(CSM in the following)
and various of their extensions have received a considerable new impetus
of interest in the last years. We mention for example the recent
preprint \cite{cardy}
where CSM are applied in the framework of conformal field theory.
From the beginning, the CSM are know to be integrable
both on the classicall and quantum levels. In the framework
of quantum systems, the distinction
between integrable and exactly solvable models can further be made,
according to the lines of Ref. \cite{turbiner}
where it was shown that quantum CSM possess both properties: 
integrable and exactly solvable. 
Off course the notion of exactly solvable operators
constitutes a particular case of Quasi-Exactly-Solvable 
(QES) operators
\cite{tur0} for which only a finite part of the spectrum can be
computed algebraically.
The possibility of having QES extensions of the
 Calogero and Sutherland Hamiltonians
 is a natural question  which was adressed
e.g. in \cite{tur2, bk}.

Recently again, it was recognized that the self-adjoint property
is not necessary for an operator $A$ to have a real spectrum. Instead
the weaker condition of invariance of $A$ under the combination
of parity $P$ and  time-reversal $T$ symmetries leads to a spectrum
which is either real or composed of pairs of complex conjugate
numbers \cite{bender}. It is therefore natural to consider
non-self adjoint but
PT-symmetric Hamiltonians describing N-particles 
quantum systems and to construct operators which
are (i) integrable or (ii) exactly solvable or (iii)
quasi-exactly solvable. Examples of such
operators were studied in  \cite{basu1}
and reconsidered more recently \cite{basu2}

Off course the problem of classifying all PT-symmetric 
(quasi)- exactly integrable operators is
 vast, one possible way to  approach it \cite{basu1,basu2}  is to
extend the well established CSM by suitably chosen non self-adjoint
but PT-symmetric extra terms and to study which algebraic
properties of the CSM are preserved or spoiled by the new terms.
In this paper we consider more general types of PT-symmetric terms
restricted by the following properties
(i) translation invariant,
(ii) first order in the derivatives,
(iii) fully symmetric under the permutations of the N
degrees of freedom.
In the three cases we put the emphasis on the (quasi) solvability
of the deformed model. We will exhibit two extensions which preserve
the exact solvability  and another one leading to a new type
of QES N-body Hamiltonian.

\section{PT-invariant Hamiltonians: rational case}
We consider the quantum N-body Hamiltonians of the form
\begin{equation}
H = H_{cal} + H_{PT}
\end{equation}
where $H_{cal}$ denotes the rational Calogero\cite{calogero}
 Hamiltonian~:
\be
   H_{cal} = 
   -\frac{1}{2} \sum_{j=1}^N \frac{\partial^2}{\partial x_j^2}
   + \frac{\omega^2}{2N}(\frac{1}{2} \tilde \tau_2 
   + \epsilon^2\sigma_1^2)
   + g \sum_{j<k} \frac{1}{(x_j-x_k)^2}
\ee
\begin{eqnarray}
& H_{PT} &= \delta\sum_{j\not=k}\frac{1}{x_j-x_k}\frac{\partial}{\partial x_j}+ \gamma\sum_{j\not=k}(x_j-x_k)\frac{\partial}{\partial x_j}\\
&        &= \delta H_0 + \gamma H_1
\end{eqnarray}

\be 
      \tilde \tau_2 \equiv \sum_{j \neq k} (x_j-x_k)^2 \ \ , \ \
\sigma_1 \equiv \sum_{j=1}^N x_j \ \ \ ,
\ee
For later convenience, we split the traditional harmonic part
into the translation-invariant piece  ($\tilde \tau_2$) and the
"center of mass coordinate" $\sigma_1$;
\be
   \sum_{j=1}^N x_j^2 = \frac{1}{N}
   (\frac{1}{2} \tilde \tau_2 + \sigma_1^2)
\ee
Setting $\epsilon=0$ the above Hamiltonian is purely translation 
invariant, the standard case is recovered for $\epsilon =1$.
The part $H_{PT}$ is not self-adjoint but symmetric under the 
$PT$ reflexion $i \rightarrow -i$ , $x_k \rightarrow - x_k$ , 
$p_k \rightarrow p_k$  $(p_k=\frac{1}{i}\frac{\partial}{\partial x_k})$,
it extends the interaction $H_0$ considered recently in 
\cite{basu2}.
In order to reveal the solvability property of $H$ we first perform
the standard  (often called ``gauge") transformation
\be
    \tilde H = \Psi_{gr}^{-1} H \Psi_{gr}  \ \ , \ \ 
    \Psi_{gr} = \beta^{\nu} E
\ee
with 
\be
       \beta = \prod_{j<k}(x_j-x_k) \ \ , \ \ 
E = \exp(-\frac{\alpha}{4} \tilde \tau_2 -
\frac{\alpha'}{2}\sigma_1^2)
\ee
After some algebra, the new Hamiltonian $\tilde H$ is obtained.
The parameters $\nu, \alpha, \alpha'$ entering in $\Psi_{gr}$
are fixed according to 
\be
   g = \nu^2 - \nu(1+2 \delta) \ \ , \ \ 
   \omega^2 = \alpha^2 N + 2 \alpha \gamma N  \ \ , \ \
    \omega = \frac {\alpha' N}{\epsilon}
\ee
in such a way that the terms with the highest power in $x_i$
as well as the most singular terms in $x_i$ disappear in $\tilde H$.
The condition on $g,\nu, \delta$ is just identical as in the case
of Ref.\cite{basu1} and lead to the same restrictions on these
constants for the eigenvectors to be nonsingular~: (i) 
$\delta > - \frac{1}{2}$ , $ 0 > g > - (\delta + \frac{1}{2})^2$
and $g>0$ for arbitrary values of $\delta$.

\subsection{Exact solvability of $H$}
The next step consists in writing $\tilde H$ in terms 
of variables which encodes the  full symmetry 
of the problem under permutations of the N degrees of freedom. 
We use here the variables introduced
in   \cite{turbiner} which we recall for completeness~:
\begin{eqnarray}
&\sigma_1 (x) &= x_1 + x_2 + \dots + x_N \\
&\sigma_2 (x) &= x_1 x_2 + x_1 x_3 + \dots + x_{N-1} x_N \\
&\sigma_3 (x) &= \sum_{i<j<k} x_i x_j x_k  \ \ \ , \ \dots \ \ \  \\
&\sigma_N (x) &= x_1 x_2 x_3 \dots x_N
\end{eqnarray}
and for $k = 2,3,\dots ,N$ we further define the 
translation invariant variables
\be
    \tau_k(x) = \sigma_k(y)\ \ \, \ \ y_j = x_j - \frac{1} {N}\sigma_1
\ee
In particular, we have
$ \tilde \tau_2 = -4N\tau_2$.
 The Laplacian operator once expressed in terms
 of these new variables reads \cite{turbiner}
\begin{eqnarray}
&\Delta &= \sum_{j=1}^N \frac{\partial^2}{\partial x_j^2}\\
&     &= N \frac{ \partial^2}{\partial\sigma_1^2}+\sum_{j,k=2}^N                  A_{jk}\frac{\partial^2}{\partial\tau_j\partial\tau_k}+                   \sum_{j=2}^N B_j\frac{\partial}{\partial\tau_j}\\               &A_{jk} &={ \frac{(N-j+1)(k-1)}{N}\tau_{j-1}\tau_{k-1}
          +\sum_{l\geq max( 1, k-j)}^N(k-j-2l)\tau_{j+l-1}\tau_{k-l-1}}\\
&-B_j &= \frac{(N-j+2)(N-j+1)}{N}\tau_{j-2}                 
\end{eqnarray} 
where the following conventions have to be understood
 \be
 \tau_0 = 1 \ \,\ \ 
 \tau_1 = 0 \ \,\ \ 
 \tau_{-p} = 0 \ \,\ \
 \tau_{N+p} = 0   \ \ {\rm for} \ p > 0
 \ee
 The PT invariant interactions $H_0, H_1$ defined in Eq.(3) can also be expressed in terms of the new variables by means of the
 following   identities~:
\begin{eqnarray}
& D &= \sum_{j=1}^N x_j \frac{\partial}{\partial x_j}\\
&   &= \sigma_1\frac{\partial}{\partial\sigma_1} + \sum_{k=2}^N k \tau_k     \frac{\partial}{\partial\tau_k}\\
&    &\equiv \sigma_1\frac{\partial}{\partial\sigma_1}+ \tilde D\\
& H_0 &=-\frac{1}{2}\sum_{j=2}^N(N-j+2)(N-j+1)
        \tau_{j-2}\frac{\partial}{\partial\tau_j}\\
& H_1 &= N\tilde D
\end{eqnarray}
 The final form of the operator $\tilde H $ reads then 
\begin{eqnarray}
\label{htilde}
& \tilde H &= \frac{-1}{2}\Delta + (\nu - \delta) H_0\nonumber \\
&             &+ \omega\sqrt{1+ \frac{\gamma^2 N^2}{\omega^2}}\tilde D
              + \omega\epsilon\sigma_1\frac{\partial}{\partial\sigma_1}
              + E_0\\
& E_0      &= \alpha\frac{N(N-1)}{2}(\nu N + 1)+ \frac{N\alpha'}{2}
           + \nu\gamma\frac{N^2(N-1)}{2}- \delta\alpha\frac{N^2(N-1)}{2} \end{eqnarray}
The solvability of the operator $ \tilde H $ (and therefore of $H$ in (1))
is revealed by the observation \cite{turbiner}
that this operator (25) preserves
the infinite flag of vector space
\be
\label{vspace}
   {\cal V}_n = {\rm span}(\sigma_1^{n_1}\tau_2^{n_2}\dots\tau_N^{n_N}\\,
   \sum_{k=1}^N
            n_k\leq n)\\, \ n \in \N
\ee
The eigenvectors of $ \tilde H $ can      therefore
be constructed by considering the restriction of
$ \tilde H $ to ${\cal V}_n $. As far as the construction
of the eigenvalues is concerned,
 it can be realized that the restriction of
$ \tilde H $ to $ {\cal V}_n $ leads  effectively to a
lower triangular matrix, the diagonal elements of which are issued
from the matrix elements  corresponding to the second line
in Eq.(\ref{htilde}).  Off course the triangularity of this
matrix is strongly related to the order adopted to enumerate
the basis elements $\tau_2^{n_2} \dots \tau_N^{n_N}$.
The relevant order can be set as follow
\begin{eqnarray}
   &0\leq  &n_N \leq n \nonumber \ ,  \\
   &0 \leq &n_{N-1} \leq n - n_{N} , \dots \ \ , \nonumber \\
   &0 \leq &n_2 \leq n - n_N - n_{N-1} \dots -n_3
\end{eqnarray}

The eigenvalues can therefore be determined
in terms of the diagonal matrix elements by using the monomials
defined in Eq.(\ref{vspace}) as a basis. 
The spectrum reads
\be
   E_{n_1, n_2, \dots , n_N} = E_0 + \omega \epsilon n_1
      + \omega \sqrt{1 + \frac{\gamma^2 N^2}{\omega^2}}
      (2 n_2 + 3 n_3 + \dots N n_N)
\ee
The spectrum of the Calogero model is off course recovered with
all its degeneracies in the limit $\gamma = 0$; the term $H_0$
does not affect these degeneracies.
However for
$\gamma \neq 0$ several degeneracies, but not all, are lifted
by the $H_1$ interaction. The eigenstates which remain degenerate
are those with $\sum_{k=2}^N k n_k = \sum_{k=2}^N k n'_k$.

\subsection{Quasi-exactly Solvable case}
In this section we consider the Hamiltonian
\be
   H = H_{cal}+\theta\sum_{i\not=j}{(x_i-x_j)}^3
      \frac{\partial}{\partial x_i}+ \tilde\theta\sum_{i\not=j}{(x_i-x_j)}^4
\ee
where $ H_{cal}$ is supplemented by a PT-invariant term
(with coupling constant $\theta$) as well as a new term of the potential.
This term is anharmonic, being of degree 4 in the position,
 and characterized by a coupling constant $\tilde\theta$.
Performing the change of function by means of $ \psi_{gr}$ leads to
\begin{eqnarray}
& \tilde H &= -\frac{1}{2}\Delta + \nu H_0 + \omega D\nonumber \\
&           &+ (\tilde\theta- \frac{\theta \alpha}{2})
\sum_{i\not=j}{(x_i-x_j)}^4 + 4N\tau_2(\frac{\alpha^2 N}{4}-\frac{\omega^2}{4N}- \theta\nu - \frac{\theta N}{4} \tilde D)
\end{eqnarray}
If the new coupling constant $ \theta, \tilde\theta $ are chosen
in such a way that
\be
\label{condition}
   \tilde\theta = \frac{\theta \alpha}{2} \ \ ,\ \
   \frac{1}{\theta}
   (\alpha^2-\frac{\omega^2}{N^2}-\frac{4\theta\nu}{N})= m \in \N
   \ , \ (\leftrightarrow \  \alpha^2 =
   \frac{\omega^2}{N^2} + 4 \frac{\theta \nu}{N} + m \theta  )
\ee
then the operator $ \tilde H $ preserves a finite dimensional vector
space which we call $ \tilde {\cal V}_m $.
As so, it is quasi-exactly solvable \cite{tur0}.
The vector space $ \tilde {\cal V}_m $ is characterized by the condition
\be
    \tau_2(m-\tilde D)\tilde{\cal V}_m =
    \tau_2(m- \sum_{k=2}^N k\tau_k \frac{\partial}{\partial\tau_k})
    \tilde{\cal V}_m\subset \tilde{\cal V}_m
\ee
A little algebra shows that $ m $ can not be arbitrary. As an example, if $ N=3 $ the vector space $ \tilde {\cal V}_m $ has the form
\be
   \tilde{\cal V}_m = {\cal P}(\frac{m}{2})\oplus
   \tau_3^2 {\cal P}(\frac{m}{2}- 3)\oplus
   \tau_3^4 {\cal P}(\frac{m}{2}- 6)\oplus\dots
\ee
if $ m $ is even $ (m\geq 2)$ and
\be
   \tilde{\cal V}_m = \tau_3 {\cal P}(\frac{m-3}{2})
                \oplus \tau_3^3{\cal P}(\frac{m-9}{2})
                \oplus \tau_3^5{\cal P}(\frac{m-15}{2})\oplus \dots
\ee
if $ m $ is odd $ ( m\geq 5)$. 
The sum off course runs with $k$ as long as $(m-6k)/2$
(resp. $(m-3-6k)/2$)
is positive for $m$ even (resp. odd).
Remark that  we have the
following inclusions
\be
 \tilde{\cal V}_m  \subset {\cal V}_{m/2} \ \ \ ,
 \tilde{\cal V}_m  \subset {\cal V}_{(m-3)/2}
\ee
and that the vector spaces $\tilde{\cal V}_m$ are preserved separately
by the operators $\Delta, H_0, D$.
 In order to illustrate these results,
we constructed the eigenvalues of the Hamiltonian in the case $N=3$,
$m=6$, for which the vector space $\tilde {\cal V}_6$ has five
dimensions.The coupling constant $\tilde\theta$ of the 
quartic interaction in (30) 
is then adjusted in terms of $ \theta, \omega, \nu$ and m by 
Eq (32).Note that Changing the value of $m$ results in 
another value for $\alpha$ and therefore
another value for $\tilde\theta$.
    The eigenvalues $E$ cannot be computed  explicitely
in terms of the coupling constants $\nu, \omega , \theta$ but,
considering the PT-symmetric term as a perturbation of the
Calogero Hamiltonian leads to the following result
\begin{eqnarray}
& E_{0,0} &= 0 + 27 (\nu-1) \frac{\theta}{\omega} + O(\theta^2) 
\nonumber \\
&E_{1,0}  &= 2\omega + 9 (\nu-5) \frac{\theta}{\omega}+ O(\theta^2) 
\nonumber \\
&E_{2,0} &= 4\omega - 9 (\nu+1) \frac{\theta}{\omega}+ O(\theta^2)
\nonumber \\
&E_{3,0} &= 6\omega - 27(\nu-3) \frac{\theta}{\omega}+ O(\theta^2) 
\nonumber \\
&E_{0,2}  &= 6\omega
\end{eqnarray}
Note that the only degenerate energy levels which occur
in the diagonalisation of the restricted to ${\cal V}_3$
corresponding to $m_2=3, m_3=0$ and $m_2=0, m_3=2$.
These vectors are included in the subspace $\tilde{\cal V}_6 $ and,
again, this new term lift the degeneracy.
The six extra vectors corresponding to ${\cal V}_3$ are not 
accessed algebraically in the presence of the new term.
We also solve numerically
the eigenvalue equation for generic values of $\nu,\theta$
and found that the spectrum remain real.
\section{PT-invariant Hamiltonians: trigonometric case}
 Here, we consider an extension of the Sutherland Hamiltonian 
 $H_{su}$ in the form
\begin{eqnarray}
\label{hfull}
& H &= H_{su} + H_{PT} + V \\
& H_{su} &= -\frac{1}{2}\Delta +                    
\frac{g}{4}\sum_{i<j}\frac{1}{\sin^2(\frac{x_i-x_j}{2})}
\end{eqnarray}
where the extra $PT$ symmetric part is chosen of the form
\begin{eqnarray}
\label{hpt2}
&H_{PT} &= \delta H_0 + \gamma H_1 \\
&       &= \delta\sum_{i<j}\cot(\frac{x_i-x_j}{2})
(\frac{\partial}{\partial x_i}-\frac{\partial}{\partial x_j})\nonumber \\
&       &+ \gamma\sum_{i<j}\sin(x_i-x_j)
(\frac{\partial}{\partial x_i}-\frac{\partial}{\partial x_j})
\end{eqnarray}
which provides a natural generalisation of Eq.(3) 
to the trigonometric case.
The Sutherland inverse sine-square potential is also supplemented
by the following terms
\begin{eqnarray}
\label{potentiel}
&V &= \theta_1\sum_{i\not=j\not=k}\cot(\frac{x_i-x_j}{2})
\cot(\frac{x_i-x_k}{2}) \nonumber \\
&  &+ \theta_2\sum_{i<j}\cos^2(\frac{x_i-x_j}{2})\nonumber \\
&  &+ \theta_3\sum_{i\not=j\not=k}\cos(\frac{x_i-x_j}{2})\cos(\frac{x_i+x_j-2 x_k}{2})
\end{eqnarray}
 where the $\theta_2$-term involves two
 body-interactions (like the original Sutherland potential)
 while the $\theta_1$ and $\theta_3$-terms involve three
 body-interactions.
Similarly the previous case, it is convenient to 
``gauge rotate"
of $H$ in (38) by using the ground state $\psi_{gr}$ of the 
Sutherland model i.e
\be
 \psi_{gr}= (\prod_{j<k}\sin(\frac{x_j-x_k}{2}))^{\nu}\ \ \,
 \ \ \ \nu^2 + \nu=g
\ee
\newpage
 In this purpose, the following identities are usefull
\begin{eqnarray}
&\beta^{-1}H_0\beta &= H_0 + \sum_{i<j}\frac{1}
{\sin^2(\frac{x_i-x_j}{2})}-\frac{N(N-1)}{2}\nonumber \\
&   &+\sum_{i\not=j\not=k}\cot(\frac{x_i-x_j}{2})
\cot(\frac{x_i-x_k}{2})\\
&\beta^{-1}H_1\beta &= H_1 + \sum_{i<j}\cos^2(\frac{x_i-x_j}{2}) 
+  \sum_{i\not=j\not=k}
\cos(\frac{x_i-x_j}{2})\cos(\frac{x_i+x_j-2x_k}{2})
\end{eqnarray}

The gauge rotation of the full Hamiltonian   (\ref{hfull})
by $\psi_{gr}$ leads to the following equivalent operator
\begin{eqnarray}
\label{suth}
&h &= -2\beta^{-\nu}H\beta^{\nu}\\
&  &= \Delta + (\nu-2\delta)H_0 + (-2\gamma)H_1 + 
E_0
\end{eqnarray}
where $E_0 \equiv \nu^2 N (N^2-1)/12$ denotes the energy
of the ground state.
In order to obtain the above formula, the various 
coupling constants are choosen according to
\be
  g = \nu(\nu-1) +\delta \ \ \ , \ \ \theta_1=-\delta \ \ \ , \ \     
  \theta_2=\theta_3= -\gamma
\ee 
in such a way that the singular terms and the non derivative 
parts of the operator $h$ cancel.
 \subsection{Exact solvability}
 The next step to reveal the solvability of the Hamiltonian
 (\ref{suth}) is to use the following change of variables
\begin{eqnarray}
&\xi_N &= e^{ix_1} + e^{ix_2}+ \dots + e^{ix_N}\\
&\eta_1 &= e^{iy_1} + e^{iy_2}+ \dots + e^{iy_N}\\
&\eta_2 &= e^{i(y_1+y_2)} + e^{i(y_1+y_3)}+ \dots + e^{i(y_{N-1}+ y_N)}\\
&\dots & \nonumber \\
&\eta_{N-1} &= \eta_1^* \ \ \ , \ \ y_i = x_i - \frac{1}{N}\sum_{k=1}^N x_k
\end{eqnarray}
The different operators can then be expressed in terms of the new variables namely
\begin{eqnarray}
&\Delta &= -N(\xi_N\frac{\partial}{\partial\xi_N})^2 - \sum_{j,k =1}^{N-1}A_{jk}\frac{\partial^2}{\partial\eta_j\partial\eta_k} - \frac{1}{N}\sum_{l=1}^{N-1}l(N-1)\eta_l\frac{\partial}{\partial\eta_l}\\
&A_{jk} &= \frac{k(N-j)}{N}\eta_j\eta_k + \sum_{l\geq max(1,k-j)}(k-j-2l)\eta_{j+l}\eta_{k-l}\\
&H_0 &= -\sum_{l=1}^{N-1}l(N-l)\eta_l\frac{\partial}{\partial\eta_l}
\end{eqnarray}

Considering first the $H_0$ interaction only 
($\gamma = \theta_2 = \theta_3 = 0$),
we observe that this PT-symmetric term
does not spoil the solvability of the Sutherland Hamiltonian
\cite{turbiner}
because both $\Delta$, $H_0$ preserve the infinite flag 
of vector spaces
 ${\cal V}_n(\eta)$ (see (\ref{vspace}) for the definition)
 as easily seen from Eq. (\ref{suth}).
It is, however,
worth to stress that the term $H_0$ leads to solvability
only if the extra three-body potential corresponding to the
$\theta_1$-term in Eq. (\ref{potentiel}) is supplemented 
to the original two-body Sutherland potential.
This contrasts with the rational
case where the addition of the term $H_0$ to the Calogero Hamiltonian
leads to another solvable model without any extra pieces in the
potential.
The interaction $H_1$ is investigated in the next section.

\subsection{Quasi Exact solvability}
We could not evaluate $H_1$ in terms of $\eta$ for generic cases
of  N but, for N=2,3, we find respectively
\begin{eqnarray}
&H_1 &= \frac{1}{2}(\eta_1^2-4)\eta_1\frac{\partial}{\partial\eta_1}\ \ ,
 \ \ \eta_1=2 \cos\frac{x_1-x_2}{2}\\
 \label{n3}
&H_1 &= \frac{1}{2}(\eta_1^2\eta_2-2\eta_2^2-3\eta_1)
\frac{\partial}{\partial\eta_1} + \frac{1}{2}
(\eta_2^2\eta_1-2\eta_1^2-3\eta_2)\frac{\partial}{\partial\eta_2}\\
&\eta_1 &= e^{iy_1} + e^{iy_2} + e^{iy_3} \ \ \ , \ \ \eta_2=\eta_1^*
\end{eqnarray}
We see that, for
 $\gamma\not=0$, the Hamiltonian $h$ is likely nor
 solvable neither quasi-exactly solvable as
 suggested by the form of $H_1$ for N=3, (see Eq(\ref{n3})).
 This operator obviously does not preserve any of the
 ${\cal V}_n(\eta)$ (see (\ref{vspace})).

For N=2, the operator (\ref{hfull}) is nevertheless QES
provided  $\theta_2=-2\gamma n , n \in \N$ in (\ref{potentiel}).
Indeed, the ``gauge rotated" operator reads in this case
\be
    h  = (\frac{1}{2}\eta_1^2 - 2) \frac{\partial^2}{\partial\eta_1^2}
    + (\frac{1}{2} + \nu - \delta)\eta_1\frac{\partial}{\partial\eta_1}
+ \frac{1}{2}\gamma(\eta_1^2-4)\eta_1\frac{\partial}{\partial\eta_1}
- \frac{1}{2} \gamma n \eta_1^2
\ee
which can be easily checked to preserve the vector space
\be
  {\cal P}_n = {\rm span}(\eta_1^n \ , \ \eta_1^{n-2} \ ,
  \ \eta_1^{n-4}\,\dots)\subset{\cal V}_n
\ee
Accordingly,  $(n+2)/2$ (resp. $(n+1)/2)$ eigenvectors
can be constructed algebraically for $n$ even (resp. odd).

Once more we notice that, in spite of the fact that (\ref{hpt2})
constitutes the natural counterpart of Eq. (3)
for the trigonometric case,
the solvability is preserved only for the interaction
$H_0$;  in contrast to the rational case, the Hamiltonian
$H_1$ spoils the solvability of the model for $N > 2$.

\section{Conclusion}
We have considered several extensions of the Calogero
and Sutherland
models by means of PT-symmetric terms which are translation invariant,
completely symmetric
and involve first derivatives.
The algebraic structure of these extended Hamiltonians is
revealed by (i) the change of function involving a common ground
state function $\psi_{gr}$, (ii) the change to appropriate symmetric
coordinates and(iii) the construction of some
vector spaces of polynomials preserved by these operators.
The completely solvable character
of the Calogero Hamiltonians is preserved for the terms called
$H_0$ and $H_1$. Another choice of interaction, involving  a
PT-symmetric term supplemented by an extra anharmonic potential,
 leads to a new Quasi-Exactly-Solvable extension of the Calogero
 model.
Note that this QES extension differs from the one obtained
first in \cite{tur2} where a potential depending
 of the variable $\tau_2$ was added.\\
Applying the  same kind of methods to the  trigonometric
(or Sutherland) model reveals that some PT-invariant terms lead
to  completely solvable extended models
only provided a suitable
extra three-body interaction is added to the potential.
For another
natural PT-symmetric term, the solvability is spoiled
 even at the price of adding new terms to the
 Sutherland original potential.
Only for $N=2$ the corresponding operator turns out to be
quasi-exactly solvable.

{\bf Acknowledgements}\\
A. N. is supported by a grant of the C.U.D..

\end{document}